\documentclass[preprint,showkeys]{revtex4}

\usepackage{amsmath}
\usepackage{graphicx}

\begin{document}

\title{Exact analysis of disentanglement for continuous variable systems and application to a two-body
system at zero temperature in an arbitrary heat bath}
\author{G. W. Ford}
\affiliation{Department of Physics, University of Michigan, Ann Arbor, MI 48109-1040 USA}
\author{R. F. O'Connell\footnote{Corresponding Author \\
E-mail: oconnell@phys.lsu.edu}}
\affiliation{Department of Physics and Astronomy, Louisiana State University, Baton
Rouge, LA 70803-4001 USA}
\date{\today}

\begin{abstract}
We outline an exact approach to decoherence and entanglement problems for
continuous variable systems. The method is based on a construction of
quantum distribution functions introduced by Ford and Lewis \cite{ford86} in
which a system in thermal equilibrium is placed in an initial state by a
measurement and then sampled by subsequent measurements. With the Langevin
equation describing quantum Brownian motion, this method has proved to be a
powerful tool for discussing such problems. After reviewing our previous work on decoherence and our recent work on disentanglement, we apply the method to the
problem of a pair of particles in a correlated Gaussian state. The initial
state and its time development are explicitly exhibited. For a single relaxation time bath at zero temperature exact
numerical results are given. The criterion of
Duan et al. \cite{duan00} for such states is used to prove that the state is
initially entangled and becomes separable after
a finite time (entanglement sudden death).
\end{abstract}

\keywords{Disentanglement; Heat Bath; quantum Langevin equation; non-Markovian; Quantum Brownian motion}

\maketitle

\section{Introduction}

Simple quantum systems do not exist in isolation but are subject to
enviromental effects which can be simple temperature (in the case of negligble dissipation) and quantum effects or,
more generally, also dissipative effects. The first quantative treatment of
such effects goes back to the phenomological equations of Bloch for the
description of nuclear magnetic resonance, with the well-known relaxation times $T_{1}$ and $T_{2}$
, which were later shown to arrive from a solution of the master equation
for a two-level system (describing a spin 1/2 system).

There has been much interest in recent years in small quantum systems,
particularly those in quantum superposition states, which are generally
entangled. Thus, in particular, for a single quantum particle in a
superposition state in a heat bath, very short decoherence times arise. In
order to calculate such times accurately, many different techniques have
been proposed such as the Feynman-Vernon functional integral approach, the
use of master equations and various stochastic methods. We have found that
the quantum Langevin equation, supplemented by use of the Wigner
distribution function, provides the basis of a powerful and physically
transparent approach to such problems. In addition, our techniques are
generally exact and lend themselves naturally to the incorporation of what
we regard as the correct initial conditions. Moreover, memory and non-Markovian effects
are naturally incorporated \cite{ford88}. 

First, we want to lay to rest the notion that there is a useful master equation.  We commence by examining initial conditions within the framework of master
equations where one starts with an initially uncoupled quantum state, a free
particle, say. Thus, the free particle is essentially at zero temperature
with no cognizance of even the zero-point oscillations of the
electromagnetic field. In addition, the initial state of the heat bath is in
equilibrium at some temperature $T$ but not coupled to the free particle.
Next, the free particle and heat bath are brought into contact and, as we
have shown explicitly \cite{ford01PRD}, the free particle receives an
initial impulse with the result that the center of the wave packet drifts to
the origin. But, since for a free particle the origin cannot be a special
point, we see that the translational invariance of the problem is broken by
the assumption that the initial state corresponds to an uncoupled system.
This problem exists in so-called "exact" master equation formulations, which
are exact only in the sense that they incorporate time-dependent
coefficients but they suffer from the same defects as the more conventional
master equations; in fact, the same results arise more easily from the use
of the initial value Langevin equation which enabled us to obtain solutions
of these "exact" master equations in a much more simplified form than one
finds in the literature \cite{ford01PRD}.

The problem with choosing an initial state corresponding to a particle at
temperature $T=0$ does not give problems with nuclear magentic resonance calculations where one
deals with relatively long times, in contrast to the short times involved in
decoherence and disentanglement calculations. However, even for the latter
cases, we have shown \cite{ford01PRD} that one could circumvent this problem
by choosing an initial corresponding to a wave packet at temperature $T$
(obtained by averaging the initial Wigner function over a thermal
distribution of initial velocities); as a result, the variance for very
short times includes the thermal spreading but the initial impulse,
resulting from bringing the quantum particle into contact with the heat
bath, still remains. The end result is that "- - worthwhile results [for the
exact master equation] can only be obtained in the high temperature limit" but, as we showed in a follow-up paper, in addition to irremedial divergencies due to zero-point fluctuations arising with exact master equations at low temperatures, in the high temperature regime (where, by convention, zero point fluctuations are neglected), problems also exist, notably the fact that the density matrix is not necessarily positive \cite{ford05}.  Moreover, in earlier work \cite{karrlein97}, Karrlein and Grabert showed in general that "- - there is no Liouville operator independent of the initial preparation - -.", which "- - is intimately connected with the failure of the Onsager regression hypothesis \cite{ford96} in the quantum regime."  In other words, there is \underline{no unique master equation}.

Turning now to our quantum Langevin equation approach \cite{ford88}, which
stemmed from the special FKM model \cite{ford65}, we considered a quantum
particle coupled to a linear passive heat bath. In the distant past the
quantum particle and heat bath are assumed to be in thermal equilibrium.
Thus, we start with a complete system that is entangled with the bath at all
times. At an initial time a "measurement" is made which prepares the system
in an initial state. Then, at a later time, a second measurement is made
which samples the state at that time. This formulation was explained in
detail with a number of applications in a long paper in Phys. Rev. A \cite
{ford07}, to which we refer the reader for further detail. In essence, we
have extended the work Ford and Lewis \cite{ford86} which itself is a quantum
extension of the work of \emph{\ }Wang-Uhlenbeck and Kolomogorov on joint
probability distributions describing a classical stochastic process. In
particular, we showed how the prescription can be extended in the form of a
general formula for the Wigner function of a Brownian particle entangled
with a heat bath. The Wigner function provides the same information as the
corresponding density matrix while making the calculations simpler and more
transparent. This enabled us to calculate decoherence times for a variety of
physical systems.

Entanglement is a subject of much current interest because of its key role
in most applications in quantum information systems \cite
{knight05,horodecki09}. Thus, its possible loss due to "entanglement sudden
death" at a finite time has led to widespread interest in investigating this
phenomenon \cite{eberly07}. The b\^{e}te noire of entangled systems is the
presence of a heat bath and its effect has generally been investigated using
master equation techniques, despite their inherent limitations \cite
{ford05,karrlein97,ford07}, as we have already pointed out. However, for the case of entangled continuous variable
states, an exact analysis is possible, as we will now show. In particular,
such systems are of interest in connection with linear optical quantum
computing.

We consider two particles, each of mass $m$, at positions $x_{1}$ and $x_{2}$
in an initially entangled Gaussian state. In the absence of a heat bath, we
already showed that this proved to be a very useful system for judging
results from the use of various entanglement measures \cite{ford10}. As we will now show, entanglement sudden death even occurs at zero temperatures.
Here, we use an \underline{exact} general prescription for treating both
decoherence decay and entanglement decay for a broad class of entangled
systems, in an \underline{arbitrary heat bath}, just as we did for the decay
of coherence of a single quantum system \cite{ford01PRA}.

For a two-particle entangled system in a heat bath, the procedure is a
straightforward generalization of the method described in \cite{ford07} for
the case of a single particle in an arbitrary heat bath so that, instead of
starting with an initial state described by a single particle Gaussian wave
function, we start with a two-particle Gaussian wave function. Then, in
order to test for separability we use the Duan et al. criterion for such
Gaussian states\cite{duan00}. In Sec. II, we present our calculation and our
conclusions are discussed in Sec. III.

\section{Gaussian state for an entangled two-particle system}

The Wigner characteristic function (the
Fourier transform of the Wigner function) is given by the obvious
generalization of Eq. (6.5) of \cite{ford07} 
\begin{equation}
\mathcal{\tilde{W}}(Q_{1},P_{1};Q_{2},P_{2};t)=\frac{\left\langle f^{\dag
}(1)e^{-i(x_{1}(t)P_{1}+m\dot{x}_{1}(t)Q_{1}+x_{2}(t)P_{2}+m\dot{x}
_{2}(t)Q_{2})/\hbar }f(1)\right\rangle }{\left\langle f^{\dag
}(1)f(1)\right\rangle },  \label{exact1}
\end{equation}
where the initial measurement is described by 
\begin{equation}
f(1)=f(x_{1}(0)-x_{1},x_{2}(0)-x_{2}).  \label{exact2}
\end{equation}
in which $f\left( x_{1},x_{2}\right) $ is the \emph{c}-number function
describing the initial measurement where $x_{1}(t)$ and $x_{2}(t)$ are the
time-dependent Heisenberg operators corresponding to the displacement of
either particle: 
\begin{equation}
x_{j}(t)=e^{iHt/\hbar }x_{j}(0)e^{-iHt/\hbar }  \label{exact3}
\end{equation}
and the brackets indicate expectation with respect to the state of the
system in equilibrium at temperature $T$.

In order to evaluate this formula we make the key assumption that particles
are linear oscillators coupled to a linear passive heat bath and that within
the bath the particles are \emph{widely separated} so that we may ignore
bath-induced interactions (a requirement imposed by most investigators, for
example \cite{qasimi08}). Thus, $x_{1}(t)$ and $x_{2}(t)$ independently
undergo quantum Brownian motion. If we repeat the discussion leading to Eq.
(6.43) of \cite{ford07} we obtain 
\begin{eqnarray}
&&\left\langle f^{\dag }(1)e^{-i(x_{1}(t)P_{1}+m\dot{x}
_{1}(t)Q_{1}+x_{2}(t)P_{2}+m\dot{x}_{2}(t)Q_{2})/\hbar }f(1)\right\rangle  
\notag \\
&=&\exp \{-\sum_{n=1}^{2}\frac{\left\langle x^{2}\right\rangle
(P_{n}^{2}-K_{n}^{2})+m^{2}\left\langle \dot{x}^{2}\right\rangle Q_{n}^{2}}{
2\hbar ^{2}}\}  \notag \\
&&\times \int_{-\infty }^{\infty }dx_{1}^{\prime }\int_{-\infty }^{\infty
}dx_{2}^{\prime }f^{\dag }(x_{1}^{\prime }+\frac{L_{1}}{2},x_{2}^{\prime }+
\frac{L_{2}}{2})f(x_{1}^{\prime }-\frac{L_{1}}{2},x_{2}^{\prime }-\frac{L_{2}
}{2})  \notag \\
&&\times \frac{1}{2\pi \sqrt{\left\langle x^{2}\right\rangle \left\langle
x^{2}\right\rangle }}\exp \{-\sum_{n=1}^{2}\frac{(x_{n}+x_{n}^{\prime })^{2}
}{2\left\langle x^{2}\right\rangle }-i(x_{n}+x_{n}^{\prime })\frac{K_{n}}{
\hbar }\},  \label{exact4}
\end{eqnarray}
where $\left\langle x^{2}\right\rangle $ and $\left\langle \dot{x}
^{2}\right\rangle $ are the equilibrium variances for displacement and
velocity, the same for each particle, and we have introduced 
\begin{equation}
K_{n}=\frac{cP_{n}+m\dot{c}Q_{n}}{\left\langle x^{2}\right\rangle }
,~~~L_{n}=GP_{n}+m\dot{G}Q_{n}.  \label{exact5}
\end{equation}
Here $G=G(t)$ is the Green function where $G(t)=\left[x(0), x(t)\right]/i\hbar$. For explicit expressions suitable for numerical computation of these functions, see Appendix A of \cite{ford07}. Also $c=c(t)\equiv \frac{1}{2}
\left\langle x(t)x(0)+x(0)x(t)\right\rangle $ is the correlation function,
again the same for each particle.

These expressions are valid for any measurement function. We now specialize
to the case where the initial measurement function is a Gaussian of the form 
\begin{equation}
f(x_{1},x_{2})=\frac{(a_{11}a_{22}-a_{12}^{2})^{1/4}}{\sqrt{2\pi }}\exp \{-
\frac{a_{11}x_{1}^{2}+2a_{12}x_{1}x_{2}+a_{22}x_{2}^{2}}{4}\}.
\label{exact6}
\end{equation}
Then (\ref{exact4}) becomes 
\begin{eqnarray}
&&\left\langle f^{\dag }(1)e^{-i(x_{1}(t)P_{1}+m\dot{x}
_{1}(t)Q_{1}+x_{2}(t)P_{2}+m\dot{x}_{2}(t)Q_{2})/\hbar }f(1)\right\rangle  
\notag \\
&=&\frac{\sqrt{a_{11}a_{22}-a_{12}^{2}}}{(2\pi )^{2}\sqrt{\left\langle
x^{2}\right\rangle \left\langle x^{2}\right\rangle }}\exp \{-\frac{
a_{11}L_{1}^{2}+2a_{12}L_{1}L_{2}+a_{22}L_{2}^{2}}{8}\}  \notag \\
&&\times \exp \{-\sum_{n=1}^{2}\frac{\left\langle x^{2}\right\rangle
(P_{n}^{2}-K_{n}^{2})+m^{2}\left\langle \dot{x}^{2}\right\rangle Q_{n}^{2}}{
2\hbar ^{2}}\}  \notag \\
&&\times \int_{-\infty }^{\infty }dx_{1}^{\prime }\int_{-\infty }^{\infty
}dx_{2}^{\prime }\exp \{-\frac{a_{11}x_{1}^{\prime 2}+2a_{12}x_{1}^{\prime
}x_{2}^{\prime }+a_{22}x_{2}^{\prime 2}}{2}\}  \notag \\
&&\times \exp \{-\sum_{n=1}^{2}\frac{(x_{n}+x_{n}^{\prime })^{2}}{
2\left\langle x^{2}\right\rangle }-i(x_{n}+x_{n}^{\prime })\frac{K_{n}}{
\hbar }\}.  \label{exact7}
\end{eqnarray}
The integral is standard Gaussian and we find 
\begin{eqnarray}
&&\left\langle f^{\dag }(1)e^{-i(x_{1}(t)P_{1}+m\dot{x}
_{1}(t)Q_{1}+x_{2}(t)P_{2}+m\dot{x}_{2}(t)Q_{2})/\hbar }f(1)\right\rangle  
\notag \\
&=&\frac{\sqrt{a_{11}a_{22}-a_{12}^{2}}}{2\pi \sqrt{(a_{11}\left\langle
x^{2}\right\rangle +1)(a_{22}\left\langle x^{2}\right\rangle
+1)-a_{12}^{2}\left\langle x^{2}\right\rangle \left\langle
x^{2}\right\rangle }}  \notag \\
&&\times \exp \{-\frac{a_{11}L_{1}^{2}+2a_{12}L_{1}L_{2}+a_{22}L_{2}^{2}}{8}
\}  \notag \\
&&\times \exp \left\{ -\sum_{n=1}^{2}\frac{\left\langle x^{2}\right\rangle
\left( P_{n}^{2}-K_{n}^{2}\right) +m^{2}\left\langle \dot{x}
^{2}\right\rangle Q_{n}^{2}}{2\hbar ^{2}}\right\}   \notag \\
&&\times \exp \left\{ -\frac{(a_{22}+\left\langle x^{2}\right\rangle
^{-1})K_{1}^{2}-2a_{12}K_{1}K_{2}+(a_{11}+\left\langle x^{2}\right\rangle
^{-1})K_{2}^{2}}{2\hbar ^{2}[(a_{11}+\left\langle x^{2}\right\rangle
^{-1})(a_{22}+\left\langle x^{2}\right\rangle ^{-1})-a_{12}^{2}]}\right\} ,
\label{exact8}
\end{eqnarray}
where we have chosen $x_{1}=x_{2}=0$ in order that the inital state be
centered at the origin. The Wigner characteristic function is 
\begin{eqnarray}
&&\tilde{W}(Q_{1},P_{1};Q_{2},P_{2};t)  \notag \\
&=&\exp \{-\frac{a_{11}L_{1}^{2}+2a_{12}L_{1}L_{2}+a_{22}L_{2}^{2}}{8}\} 
\notag \\
&&\times \exp \{-\sum_{n=1}^{2}\frac{\left\langle x^{2}\right\rangle
(P_{n}^{2}-K_{n}^{2})+m^{2}\left\langle \dot{x}^{2}\right\rangle Q_{n}^{2}}{
2\hbar ^{2}}\}  \notag \\
&&\times \exp \left\{ -\frac{(a_{22}+\left\langle x^{2}\right\rangle
^{-1})K_{1}^{2}-2a_{12}K_{1}K_{2}+(a_{11}+\left\langle x^{2}\right\rangle
^{-1})K_{2}^{2}}{2\hbar ^{2}[(a_{11}+\left\langle x^{2}\right\rangle
^{-1})(a_{22}+\left\langle x^{2}\right\rangle ^{-1})-a_{12}^{2}]}\right\} ,
\label{exact10}
\end{eqnarray}

This becomes simpler in the free particle limit :$\ \left\langle
x^{2}\right\rangle \rightarrow \infty $. Noting that near the center of an oscillator potential the motion is that of a free particle so that, in this limit, the measurement function \cite{karrlein97} is, in essence, the wave function for the initial state of the free particle.  In essence, $\left\langle x^{2}\right\rangle \rightarrow \infty$ corresponds to the range of the oscillator getting larger or, concomitantly, the oscillator becomes so weak as to be indistinguishable from that of a free particle. Thus, in this limit  
\begin{eqnarray}
\tilde{W}(Q_{1},P_{1};Q_{2},P_{2};t) &=&\exp \{-\frac{
a_{11}L_{1}^{2}+2a_{12}L_{1}L_{2}+a_{22}L_{2}^{2}}{8}\}  \notag \\
&&\times \exp \left\{ -\frac{
a_{22}P_{1}^{2}-2a_{12}P_{1}P_{2}+a_{11}P_{2}^{2}}{2\hbar ^{2}\left(
a_{11}a_{22}-a_{12}^{2}\right) }\right\}   \notag \\
&&\times \exp \{-\sum_{n=1}^{2}\frac{sP_{n}^{2}+m\dot{s}P_{n}Q_{n}+m^{2}
\left\langle \dot{x}^{2}\right\rangle Q_{n}^{2}}{2\hbar ^{2}}\},
\label{exact12}
\end{eqnarray}
wher $s=s(t)=\left\langle \left( x(t)-x(0)\right) ^{2}\right\rangle
=2\left\langle x^{2}\right\rangle -2c(t)$ is the mean square displacement.

Then, in particular, we find the initial state of the Wigner characteristic
function to be 
\begin{eqnarray}
\tilde{W}(Q_{1},P_{1};Q_{2},P_{2};0) &=&\exp \{-\frac{
a_{11}Q_{1}^{2}+2a_{12}Q_{1}Q_{2}+a_{22}Q_{2}^{2}}{8}\}  \notag \\
&&\times \exp \left\{ -\frac{
a_{22}P_{1}^{2}-2a_{12}P_{1}P_{2}+a_{11}P_{2}^{2}}{2\hbar ^{2}\left(
a_{11}a_{22}-a_{12}^{2}\right) }\right\}   \notag \\
&&\times \exp \{-\sum_{n=1}^{2}\frac{m^{2}\left\langle \dot{x}
^{2}\right\rangle Q_{n}^{2}}{2\hbar ^{2}}\}.  \label{exact13}
\end{eqnarray}

For simplicity, we will henceforth confine ourselves to the symmetric case
where $a_{22}=a_{11}$. In this case we can write the Wigner characteristic
function in the form 
\begin{equation}
\tilde{W}(P_{1},Q_{1},P_{2},Q_{2})=\exp \left\{ -\frac{1}{2}\mathbf{X}\cdot 
\mathbf{M}\cdot \mathbf{X}\right\} ,  \label{exact14}
\end{equation}
where 
\begin{equation}
\mathbf{X}=\left( 
\begin{array}{c}
\frac{LP_{1}}{\hbar } \\ 
\frac{Q_{2}}{\hbar } \\ 
\frac{LP_{2}}{\hbar } \\ 
\frac{Q_{2}}{L}
\end{array}
\right) ,\qquad \mathbf{M}=\left( 
\begin{array}{cccc}
G_{11} & G_{12} & C_{11} & C_{12} \\ 
G_{12} & G_{22} & C_{21} & C_{22} \\ 
C_{11} & C_{21} & G_{11} & G_{12} \\ 
C_{12} & C_{22} & G_{12} & G_{22}
\end{array}
\right) =\left( 
\begin{array}{cc}
\mathbf{G} & \mathbf{C} \\ 
\mathbf{C} & \mathbf{G}
\end{array}
\right) .  \label{exact15}
\end{equation}
Here $L$ is a constant of dimension length, introduced to make the matrix
elements of \ the correlation matrix $\mathbf{M}$ and the elements of $
\mathbf{X}$ dimensionless. For the state with the Wigner characteristic
function given by the symmetric limit of (\ref{exact12}), we find that 
\begin{eqnarray}
G_{11} &=&\frac{1}{L^{2}}\left[ \frac{a_{11}}{a_{11}^{2}-a_{12}^{2}}+\left( 
\frac{\hbar G}{2}\right) ^{2}a_{11}+s\right] ,  \notag \\
G_{12} &=&\left( \frac{\hbar G}{2}\right) \left( \frac{m\dot{G}}{2}\right)
a_{11}+\frac{m\dot{s}}{2\hbar },  \notag \\
G_{22} &=&L^{2}\left[ \frac{m^{2}\langle \dot{x}^{2}\rangle }{\hbar ^{2}}
+\left( \frac{m\dot{G}}{2}\right) ^{2}a_{11}\right] ,  \notag \\
C_{11} &=&\frac{1}{L^{2}}\left[ -\frac{a_{12}}{a_{11}^{2}-a_{12}^{2}}+\left( 
\frac{\hbar G}{2}\right) ^{2}a_{12}\right] ,  \notag \\
C_{12} &=&C_{21}=\left( \frac{\hbar G}{2}\right) \left( \frac{m\dot{G}}{2}
\right) a_{12},  \notag \\
C_{22} &=&L^{2}\left( \frac{m\dot{G}}{2}\right) ^{2}a_{12}.  \label{exact16}
\end{eqnarray}
In these expressions we recall that $G$ is the Green function and $
s=\left\langle (x(t)-x(0)^{2}\right\rangle $ is the mean square
displacement, the same for both particles \cite{ford07}. 

In order to discuss entanglement, Duan et al. \cite{duan00} perform a
sequence of rotations and squeezes to bring $\mathbf{M}$ to a form in which 
\begin{equation}
\mathbf{G}=\left( 
\begin{array}{cc}
g & 0 \\ 
0 & g
\end{array}
\right) ,\quad \mathbf{C}=\left( 
\begin{array}{cc}
c & 0 \\ 
0 & c^{\prime }
\end{array}
\right) \mathbf{.}  \label{exact18}
\end{equation}
Since the determinants are invariant under these transformations, we have
the following simple relations for determining the quantities $g$, $c$ and $
c^{\prime }$ in terms of these invariants 
\begin{equation}
\det \mathbf{G}=g^{2},\quad \det \mathbf{C}=cc^{\prime },\quad \det \mathbf{M
}=\left( g^{2}-c^{2}\right) \left( g^{2}-c^{\prime 2}\right) .
\label{exact19}
\end{equation}
The necessary and sufficient condition that the state be disentangled is
equivalent to the inequality 
\begin{equation}
\sqrt{\left( g-c\right) \left( g+c^{\prime }\right) }\geq \frac{1}{2}.
\label{exact20}
\end{equation}
This result is equivalent to that obtained by the Duan et al. analysis,
specialized to the symmetric case.

We have calculated these quantities for the case of two particles coupled to
a single-relaxation time bath \cite{ford07} at \emph{zero temperature}. This
heat bath is characterized by a memory function of the form 
\begin{equation}
\mu (t)=\frac{\zeta}{\tau}e^{-t/\tau}\theta (t),  \label{exact21}
\end{equation}
in the quantum Langevin equation \cite{ford88}. Here $\theta (t)$ is the
Heaviside function and we note that in the limit $\tau \rightarrow 0$ this
becomes the Ohmic memory function $\mu (t)=2\zeta\delta (t)$. The
corresponding Fourier transform of (\ref{exact21}) is 
\begin{equation}
\tilde{\mu}(\omega )=\frac{\zeta}{1-i\omega\tau}  \label{exact22}
\end{equation}
whose Ohmic limit is $\zeta \equiv m\gamma$.

Defining
\begin{equation}
b_{ij}=\frac{\zeta}{\hbar}a_{ij}
\end{equation}
which are dimensionless constants proportional to the $a_{ij}$, we choose $L^{2}=\hbar /\zeta$ and $\gamma /\tau = \frac{1}{5}$ for two different selections of the $b_{ij}$ quantities.  Thus, the choice $b_{11}=5$, $b_{12}=4$ implies that the ratio $(a_{12}/a_{11})=\frac{4}{5}$ whereas the choice $b_{11}=5,000$, $b_{12}=4,999$ implies that $(a_{12}/a_{11})=0.9998$.  Thus, as expected, the latter choice corresponding to relatively larger $a_{12}$, encounters sudden death at a later time.  In Fig. 1, we plot the left hand side of the above Duan inequality versus $\gamma t$.  Note that the curve crosses 0.5 (signifying entanglement sudden death) at two different $\gamma t$ values.  Thus, we see exactly that entanglement sudden death occurs later for larger $a_{12}/a_{11}$ values.

\section{conclusions}

The generic problem of a quantum system in an environment (heat bath) has been tackled by two main approaches, the Feynman-Vernon approach, with the use of master equations, and the quantum Langevin equation approach.  We have argued that the latter method is generally superior as it treats the whole system as being completely entangled in thermal equalibrium to begin with.

How the system attains thermal equilibrium is often referred to as the zeroth law of thermodynamics \cite{uhlenbeck63}, which goes back to the fundamental ideas of Boltzmann and Gibbs.  In essence, the microscopic laws are time-reversal invariant and the Poincar\'{e} recurrence theorem seems to preclude the achievement of equilibrium.  However, the latter can be achieved by recognizing that thermal equilibrium is a macroscopic notion and that the relaxation to equilibrium depends on coarse graining and also the Hamiltonian \cite{uhlenbeck63}.  In practice, as we have done in our initial paper on the quantum Langevin equation \cite{ford88} and later in \cite{ford01PRD}, in our discussion of the inhomogeneous equation (see section IV of \cite{ford88}), we have chosen the retarded solution, thereby breaking the time-reversal invariance of the original equations.  This could be achieved, for instance, by fastening the quantum particle to a large mass in the distant past so that it is held fixed at $x=0$ say with zero momentum.  The large (eventually infinite) number of oscillators are then allowed to come to equilibrium at temperature $T$, say, by weak interaction with another bath (similar to how a collection of particles in a container come to equilibrium by interacting with the walls of the container).  Then, still in the distant past, the system is released and the subsequent motion is governed by the appropriate Hamiltonian.  As we concluded in \cite{ford88}, this "is typical of the way time-reversal invariance is broken in macroscopic systems: they describe only the time development of a class of solutions of the microscopic equations."  The end result is that at $t=0$, say, our complete system is in thermal equilibrium at temperature $T$.  The system then develops unitarily in time after which we prepare the system in a desired state by means of a first measurement on the quantum particle.  Then, at a later time, we do a second measurement which tells us how the system has developed in time due to environmental effects.  If desired, subsequent measurements can be carried out in a similar manner.  It should be again emphasized that our procedure is exact.

Our method applies to arbitrary heat baths and arbitrary temperatures.  In particular, calculations at zero temperature are readily carried out \cite{ford03} without encountering the irremedial divergences associated with exact master equations \cite{ford01PRD}.  Having previously applied this method to a variety of decoherence problems (involving a single particle prepared in a variety of initial states and analyzing its subsequent development in a arbitrary heat bath at arbitrary temperatures), we concluded by writing a detailed paper \cite{ford07}, showing in particular that our work constitutes a quantum extension of the classical stochastic theory on joint probability distributions.

Next, we turned to the disentanglement problem which involves more than one particle.  In particular, for the two-particle system, we commenced by showing that our general techniques were very useful for judging disentanglement in the absence of a heat bath \cite{ford10}.  Next, in our present paper, we have extended our work in \cite{ford07} to the case of two particles in an initial correlated Gaussian state and we studied the time development of this state in an arbitrary heat bath at zero temperature.  We found that the state which is initially entangled becomes separable after a finite time.  Thus, entanglement sudden death is also prevalent in continuous variable systems, as well as the more often studied qubit systems \cite{eberly07}, which should raise concern for the designers of all entangled systems.  A key question is the dependence of the sudden death on the number of particles.  The procedure is again a generalization of the method described in \cite{ford07} but we expect that the computational task will be formidable, unless we discover some creative approaches.

\section*{ACKNOWLEDGMENT}

This work was partially supported by the National Science Foundation under
Grant No. ECCS-0757204.

\newpage

\begin{figure}
\includegraphics{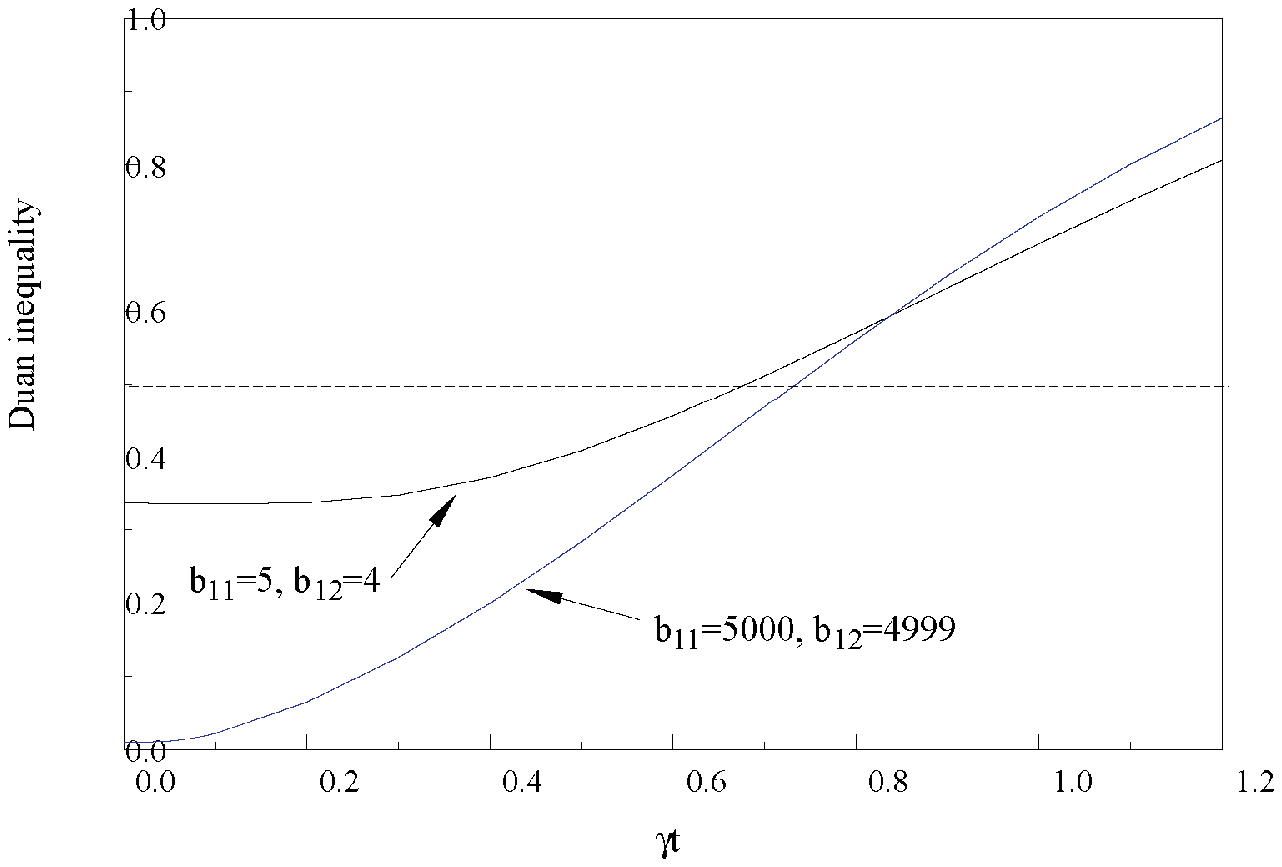}
\caption{The left side of the Duan inequality (17) as a function of time $\gamma t$, for two values of the parameters appearing in the initial wave function (6).  The horizontal dotted line signifies where entanglement sudden death occurs.}
\end{figure}

\end{document}